\documentclass[preprint, superscriptaddress, showpacs,preprintnumbers,amsmath,amssymb,pra]{revtex4}
\usepackage{graphicx}

\begin{document}

\thispagestyle{empty}

\title{Impact of graphene coating on the atom-plate interaction}

\author{
G.~L.~Klimchitskaya}
\affiliation{Central Astronomical Observatory at Pulkovo of the Russian Academy of Sciences, St.Petersburg,
196140, Russia}
\affiliation{Institute of Physics, Nanotechnology and
Telecommunications, St.Petersburg State
Polytechnical University, St.Petersburg, 195251, Russia}

\author{
V.~M.~Mostepanenko}
\affiliation{Central Astronomical Observatory at Pulkovo of the Russian Academy of Sciences, St.Petersburg,
196140, Russia}
\affiliation{Institute of Physics, Nanotechnology and
Telecommunications, St.Petersburg State
Polytechnical University, St.Petersburg, 195251, Russia}

\begin{abstract}
Using the recently proposed quantum electrodynamical formalism,
we calculate the Casimir-Polder free energies and forces between
the ground state atoms of Rb, Na, Cs and He${}^{\ast}$ and the
plates made of Au, Si, sapphire and fused silica coated with a
graphene sheet. It is shown that the graphene coating has no
effect on the Casimir-Polder interaction for metallic plates,
but influences significantly for plates made of dielectric
materials. The influence of graphene coating increases with
decreasing static dielectric permittivity of the plate material
and the characteristic frequency of an atomic dynamic polarizability.
Simple analytic expressions for the classical limit of the
Casimir-Polder free energy and force between an atom and
a graphene-coated plate are obtained. From the comparison with the
results of numerical computations, the application region of
these expressions is determined.
\end{abstract}

\maketitle

\section{Introduction}

The interaction of a polarizable atom in the ground state with an
ideal metal plate at zero temperature, caused by the zero-point
oscillations of the electromagnetic field, was described by
Casimir and Polder \cite{1}. It was generalized to the plate
materials characterized by the frequency-dependent dielectric
permittivity $\varepsilon(\omega)$ at nonzero temperature in the
framework of the Lifshitz theory \cite{2}.
In succeeding years much work has been done to investigate the
dependence of the Casimir-Polder interaction on the
characteristics
of atoms and properties of plate materials \cite{3,4,5,6,7,8}.
These theoretical investigations were stimulated by the
experimental studies of quantum reflection \cite{9,10,11} and
Bose-Einstein condensation \cite{12,13,14} of different atoms
near material surfaces. Furthermore, after a qualitative
experimental demonstration of the atom-plate interaction
\cite{15},
more precise measurement of the thermal Casimir-Polder force has
been performed \cite{16} using the advantages of Bose-Einstein
condensation. The results of this measurement were used to obtain
stronger constraints on the Yukawa-type corrections to Newtonian
gravitational law \cite{17} and on the coupling constants of
axion-like particles \cite{18}.

After the discovery of graphene, which is a two-dimensional sheet
of carbon atoms, much theoretical attention was given to
the interaction of atoms with this unusual
material. These investigations became quantitative with the
development of the Dirac model of graphene which assumes the
linear dispersion relation for the graphene bands at low
energies \cite{19,20}. First, the Dirac model was used to find
the reflection properties of electromagnetic oscillations on
graphene and calculate the van der Waals and Casimir interaction
between two graphene sheets and between a graphene sheet and a
material plate \cite{21,22,23,24,25,26,27,28,29}.
Next, the Casimir-Polder interaction of different atoms with
graphene has been calculated \cite{30,31,32,33,34}.
This was done by effectively expressing the reflection properties
of graphene either in terms of the density-density correlation
functions \cite{22,25,26,33} or via the components of the
polarization tensor in (2+1)-dimensional space-time
\cite{21,24,27,28,29,32,34}.
In so doing, it was discovered that the thermal effect in the
Casimir and Casimir-Polder interactions with graphene sheet
becomes relatively large at much shorter separations than for
usual materials \cite{22,24,25,27,28,29,32,34}.

Until the present time, the Casimir-Polder interaction between
atoms and a graphene sheet has not been measured.
This is a complicated problem bearing in mind that it would be
difficult to preserve the flatness of a freestanding
graphene under the
influence of force on the source side of atoms.
The pioneer measurement of the Casimir force in graphene research
was performed in the configuration of an Au sphere and a graphene
sheet deposited on a SiO${}_2$ film covering a Si plate \cite{35}.
 The comparison of the experimental data with the Lifshitz theory
required knowledge of the reflection coefficients of
electromagnetic
fluctuations on graphene-coated substrates. Coefficients of this
kind were found in Refs.~\cite{25,36,37}, but expressed via the
longitudinal and transverse density-density correlation functions
(or, equivalently, respective electric susceptibilities) of
graphene. However, only the longitudinal correlation function was
explicitly known and only at zero temperature, and this made
precise calculations hard. The situation has been changed after
the publication of Ref.~\cite{38} where both the longitudinal and
transverse density-density correlation functions were found at
any nonzero temperature by expressing them via the explicitly known
components of the polarization tensor. Simultaneously with this,
the reflection coefficients on graphene-coated substrates were
immediately expressed via the components of the polarization
tensor of graphene and the dielectric permittivity of a
substrate \cite{39}. When compared with the measurement data
of the experiment \cite{35}, the Lifshitz theory using these
reflection coefficients has led to a very good agreement
\cite{39}.
This opened up opportunities to the investigation of the
Casimir-Polder interaction between different atoms and
graphene-coated plates.

In this paper, we use the recently proposed formalism of
Ref.~\cite{39} to calculate the Casimir-Polder free energies
and forces between different atoms and graphene-coated
plates made of different materials. We consider the atoms
of Rb, Na, Cs and He${}^{\ast}$ (metastable helium) and the
plates made of Au, Si, sapphire (Al${}_2$O${}_3$) and fused
silica (SiO${}_2$) coated with a graphene sheet.
We determine the influence of graphene coating on the
Casimir-Polder free energy and force at room temperature.
According to our results, graphene coating of metallic plates
has no effect on the atom-plate interaction.
For dielectric plates the impact of graphene coating is the
greater, the smaller is the static dielectric permittivity of
plate material. We show that the impact of graphene coating on
the Casimir-Polder interaction for a dielectric plate increases
with decreasing characteristic frequency of the atomic dynamic
polarizability. We also find the classical  limit for the
Casimir-Polder force between atoms and graphene-coated plates.
The main terms in this limit do not depend on the plate
material, in contrast to the case of uncoated plates.

The paper is organized as follows. In Sec.~II we briefly describe
the formalism of the Lifshitz theory adapted to calculate the
Casimir-Polder interaction between atoms and graphene-coated
plates. Then the computational results for the influence of
graphene on the free energy are presented. In Sec.~III the
influence of graphene coating on the Casimir-Polder force is
investigated and the classical (high-temperature) limit is
derived. Section~IV contains our conclusions and discussion.

\section{Free energy of atoms interacting with graphene-coated
plates}

Here, we introduce the used formalism, specifically, the
reflection coefficients of the electromagnetic fluctuations
on a graphene-coated plate, where graphene is desribed by the
polarization tensor, and the plate material by the
frequency-dependent dielectric permittivity. Then we calculate
the free energy of the Casimir-Polder interaction between
different atoms and graphene-coated plates made of various
materials and determine the influence of graphene coating.

\subsection{Formalism and notations}

We consider a ground state atom described by the atomic dynamic
polarizability $\alpha(\omega)$ at a separation $a$ from the
plane plate described by the frequency-dependent dielectric
permittivity $\varepsilon(\omega)$ coated with a graphene sheet.
It is assumed that the plate is in thermal equilibrium with an
environment at temperature $T$. The free energy of the
Casimir-Polder interaction between an atom and a graphene-coated
plate is given by the Lifshitz formula \cite{2}.
For the purpose of computations it is convenient to express it
in terms of the dimensionless variables as follows \cite{40}:
\begin{eqnarray}
&&
{\cal F}_{\!\! g}(a,T)=-\frac{k_BT}{8a^3}\sum_{l=0}^{\infty}
{\vphantom{\sum}}^{\prime}
\alpha(i\zeta_l\omega_c)
\int_{\zeta_l}^{\infty}\!\!\!dye^{-y}
\label{eq1}\\
&&~
\times\left\{2y^2R_{\rm TM}(i\zeta_l,y)-\zeta_l^2\left[
R_{\rm TM}(i\zeta_l,y)+R_{\rm TE}(i\zeta_l,y)\right]
\right\}.
\nonumber
\end{eqnarray}
\noindent
Here, $k_B$ is the Boltzmann constant,
the dimensionless Matsubara frequencies $\zeta_l$ are
expressed via the dimensional ones by
$\zeta_l\equiv\xi_l/\omega_c=2a\xi_l/c$,
where $\xi_l=2\pi k_BTl/\hbar$ with
$l=0,\,1,\,2,\,\ldots$, $\omega_c=c/(2a)$,
and the prime on the summation sign indicates
that the term with $l=0$ is divided by two.
The dimensionless variable $y$ is connected
with the projection of the wave vector on the plane of
plate, $k_{\bot}$, by the equation
\begin{equation}
y=2a\left(k_{\bot}^2+\frac{\xi_l^2}{c^2}
\right)^{1/2}.
\label{eq2}
\end{equation}

The reflection coefficients $R_{\rm TM,TE}$ of the
electromagnetic fluctuations on the graphene-coated
plate for the transverse-electric (TE) and
transverse-magnetic (TM) polarizations can be expressed via the
dielectric permittivity of the plate material
\begin{equation}
\varepsilon_l\equiv
\varepsilon^{(n)}(i\xi_l)=
\varepsilon^{(n)}(i\zeta_l\omega_c),
\label{eq3}
\end{equation}
\noindent
and the dimensionless polarization tensor of graphene in
(2+1)-dimensional space-time, $\tilde{\Pi}_{kn}$
($k,\,n=0,\,1,\,2$), connected with the dimensional one,
$\Pi_{kn}$, by the equation
\begin{equation}
\tilde{\Pi}_{kn}\equiv\tilde{\Pi}_{kn}(i\zeta_l,y)=
\frac{2a}{\hbar}\Pi_{kn}.
\label{eq4}
\end{equation}
\noindent
The explicit expressions for these reflection coefficients follow
from the results of Refs.~\cite{26,36,37} if to take into account
the relationship between the density-density correlation function
and the polarization tensor found in Ref.~\cite{38}.
Alternatively, the same expressions were immediately obtained in
Ref.~\cite{39} by other means.
They are given by \cite{39}
\begin{eqnarray}
&&
R_{\rm TM}(i\zeta_l,y)=\frac{\varepsilon_ly+
k_l\left(\frac{y}{y^2-\zeta_l^2}\tilde{\Pi}_{00}-1
\right)}{\varepsilon_ly+k_l
\left(\frac{y}{y^2-\zeta_l^2}\tilde{\Pi}_{00}+1
\right)},
\nonumber \\[-1mm]
&&\phantom{aaa}
\label{eq5} \\[-2mm]
&&
R_{\rm TE}(i\zeta_l,y)=\frac{y-k_l-
\left(\tilde{\Pi}_{\rm tr}-\frac{y^2}{y^2-\zeta_l^2}
\tilde{\Pi}_{00}\right)}{y+k_l+
\left(\tilde{\Pi}_{\rm tr}-\frac{y^2}{y^2-\zeta_l^2}
\tilde{\Pi}_{00}\right)},
\nonumber
\end{eqnarray}
\noindent
where
$\tilde{\Pi}_{\rm tr}$ is the sum of the spatial components
$\tilde{\Pi}_{1}^{\,1}$ and $\tilde{\Pi}_{2}^{\,2}$,
 and the following
notation is introduced
\begin{equation}
k_l\equiv
\sqrt{y^2+(\varepsilon_l-1)\zeta_l^2}.
\label{eq6}
\end{equation}

The computations below require  analytic expressions for the
quantities $\tilde{\Pi}_{00}$ and $\tilde{\Pi}_{\rm tr}$.
These quantities depend on $T$ both explicitly, as on a parameter, and
implicitly through the Matsubara frequencies.
It was shown \cite{24,27,28,32} that an explicit dependence
on $T$ influences the computational results for the
free energy and force only through the zero-frequency
term of the Lifshitz formula $l=0$.
In the region of separations considered below ($a\geq 100\,$nm)
all terms of
Eq.~(\ref{eq1}) with $l\geq 1$ without the loss of accuracy can
be calculated using the simplified polarization tensor defined
at $T=0$ and depending on $T$ only implicitly through the
Matsubara frequencies.

We consider undoped gapless graphene (as shown in Ref.~\cite{32},
at $T=300\,$K the influence of nonzero gap below 0.1\,eV on the
computational results is neglibigly small).
Under these conditions, at $\zeta_0=0$ the
 exact expressions  for the temperature-dependent
components of the polarization tensor in
 Eq.~(\ref{eq5}) are
 the following \cite{24,27,32}:
\begin{eqnarray}
&&
\tilde{\Pi}_{00}(0,y)=\frac{8\alpha\tau}{\pi\tilde{v}_F^2}
\int_{0}^{1}\!\!dx\ln\left(2
\cosh\frac{\pi\theta}{\tau}\right),
\label{eq7} \\
&&
\tilde{\Pi}_{\rm tr}(0,y)-\tilde{\Pi}_{00}(0,y)=
8\alpha\tilde{v}_F^2y^2\!\int_{0}^{1}\!\!\!dx
\frac{x(1-x)}{\theta}
\tanh\frac{\pi\theta}{\tau}.
\nonumber
\end{eqnarray}
\noindent
Here, $\alpha=e^2/(\hbar c)$ is the fine-structure constant,
the  dimensionless temperature parameter $\tau$ is defined as
$\tau\equiv 4\pi ak_BT/(\hbar c)$,
$\tilde{v}_F=v_F/c$, where $v_F\approx 9\times 10^5\,$m/s
is the Fermi velocity in graphene \cite{41,42}, and the function
$\theta$ is given by
\begin{equation}
\theta\equiv\theta(x,y)=\tilde{v}_Fy
\sqrt{x(1-x)}.
\label{eq8}
\end{equation}

As explained above, at all Matsubara frequencies $\zeta_l$
with $l\geq 1$ we can
use the following components of the
polarization tensor found in Ref.~\cite{21} at $T=0$,
where the continuous parameter $\zeta$ is
replaced with the discrete $\zeta_l$:
\begin{eqnarray}
&&
\tilde{\Pi}_{00}(i\zeta_l,y)=\pi\alpha
\frac{y^2-\zeta_l^2}{f(\zeta_l,y)},
\label{eq9} \\
&&
\tilde{\Pi}_{\rm tr}(i\zeta_l,y)-
\frac{y^2}{y^2-\zeta_l^2}\tilde{\Pi}_{00}(i\zeta_l,y)=
\pi\alpha f(\zeta_l,y),
\nonumber
\end{eqnarray}
\noindent
where the function $f$ is defined as
\begin{equation}
f(\zeta_l,y)=[\tilde{v}_F^2y^2+
(1-\tilde{v}_F^2)\zeta_l^2]^{1/2}.
\label{eq10}
\end{equation}
\noindent
As a result, the free energy of the Casimir-Polder interaction
between an atom and a material plate coated with graphene can
be calculated using Eqs.~(\ref{eq1}) and (\ref{eq5})
supplemented by Eqs.~(\ref{eq6})--(\ref{eq10}).

\subsection{Influence of graphene on the free energy}

To find the influence of the graphene coating on the
Casimir-Polder
free energy, we calculate the free energies of atom-plate
interaction in the absence, ${\cal F}$, and in the presence,
${\cal F}_{\!\! g}$, of graphene coating [in the former case
the polarization operator in Eq.~(\ref{eq5}) should be put equal
to zero]. We begin with an atom of Rb interacting with the
plates made of Au, Si, sapphire (Al${}_2$O${}_3$), and fused
silica (SiO${}_2$). At separations above 100\,nm considered here
one can describe the atomic dynamic polarizability with
sufficient precision using the single-oscillator model
\cite{3,6}
\begin{equation}
\alpha(i\xi_l)=\frac{\alpha(0)}{1+\frac{\xi_l^2}{\omega_0^2}},
\label{eq11}
\end{equation}
\noindent
where $\alpha(0)$ is the static polarizability and $\omega_0$ is
the characteristic frequency.  For the atom of Rb one has
$\alpha(0)=319.9\,$a.u. and $\omega_0=5.46\,$eV
(note that 1\,a.u. of polarizability is equal to
$1.482\times 10^{-31}\,\mbox{m}^3$).

To perform computations using Eqs.~(\ref{eq1}) and (\ref{eq5}),
one also needs the dielectric permittivities of plate materials
$\varepsilon_l$ computed at the imaginary Matsubara frequencies.
For Au the values of $\varepsilon_l$ were obtained by means of
the Kramers-Kronig relation from the tabulated optical data
\cite{44} extrapolated down to zero frequency \cite{40,45}.
Note that for the atom-plate interaction the computational
results for the free energy and force do not depend on the type
of extrapolation of the optical data by means of the Drude or
the plasma model \cite{40,45} (the values of the plasma frequency
$\omega_p=9\,$eV and the relaxation parameter $\gamma=0.035\,$eV
have been used in extrapolations). For high-resistivity Si
the values of $\varepsilon_l$ were obtained with the help of
the Kramers-Kronig relation from
the tabulated data \cite{46}.
 As to other dielectric materials (sapphire and fused
silica), we have used sufficiently precise analytic
representations for their frequency-dependent dielectric
permittivities \cite{47}.

In Fig.~1, the ratios of the Casimir-Polder free energies
${\cal F}_{\!\! g}/{\cal F}$ computed for the interaction
of a Rb atom with Au, Si, Al${}_2$O${}_3$ and SiO${}_2$
plates at $T=300\,$K are shown by the four solid lines as
functions of separation. As can be seen in this figure,
there is no influence of graphene coating on the
Casimir-Polder interaction of Rb atom with an Au plate.
The same holds for any atom interacting with any plate made
of a metal or metallic-type semiconductor.
At the same time, according to Fig.~1, there is considerable
influence of graphene coating on the interaction free energy
of a Rb atom with plates made of different dielectric
materials. Thus, even at the shortest separation, $a=100\,$nm,
the quantity ${\cal F}_{\!\! g}/{\cal F}$ takes the values
1.011, 1.038, and 1.10 for Si, Al${}_2$O${}_3$ and SiO${}_2$
plates, respectively. At the separation of $a=200\,$nm the
respective values of ${\cal F}_{\!\! g}/{\cal F}$  are equal
to 1.015, 1.043, and 1.12, and at $a=6\,\mu$m achieve
1.19, 1.22, and 1.70, respectively.
As is seen in Fig.~1, the influence of graphene coating on the
Casimir-Polder free energy increases with decreasing static
dielectric permittivity of the plate material. The largest
influence is obtained for a SiO${}_2$ plate
($\varepsilon_0=3.8$). The influence of graphene coating
becomes weaker for Al${}_2$O${}_3$ and Si plates
($\varepsilon_0=10.1$ and 11.7, respectively).

Large influence of graphene coating on the Casimir-Polder
interaction in the case of dielectric plates is explained
by the extraordinary large thermal effect inherent in this
two-dimensional material. As an exaple, in Fig.~2(a) the
ratios of the free energy at $T=300\,$K to the energy at
$T=0\,$K for a Rb atom interacting with an uncoated
(the line 1) and graphene-coated (the line 2) SiO${}_2$
plate are presented as functions of separation.
As is seen in this figure, for a graphene-coated plate the
thermal effect is much more pronounced than for an uncoated one.
This leads to a faster increase of the line 2, compared
to the line 1, with the increase of separation (which is
equivalent to the increase of temperature in calculations
of the Casimir and Casimir-Polder forces \cite{40}).

To explain the role of the thermal effect in more detail, in
Fig.~2(b) we plot the ratios ${\cal F}_{\!\! g}/{\cal F}$
for the Casimir-Polder interactions with a graphene-coated
and an uncoated SiO${}_2$ plates as functions of separation
at $T=300\,$K (the solid line) and at $T=0\,$K (the dashed
line).
Thus, the solid line reproduces the upper line in Fig.~1,
whereas the dashed line represents the ratio of the
Casimir-Polder energies, $E_g/E$, for a graphene-coated and
uncoated SiO${}_2$ plates. As can be seen in Fig.~2(b),
large influence of graphene coating occurs only at $T=300\,$K.
For the dashed line computed at $T=0\,$K one has
$E_g/E=1.07$ and 1.03 at $a=100\,$nm and $6\,\mu$m,
respectively, i.e., the role of graphene coating is limited
to only a few persent.

Now we investigate the role of graphene coating of a plate
interacting with different atoms. For this purpose we
calculate the quantity ${\cal F}_{\!\! g}/{\cal F}$ as a
function of separation for atoms of Rb, Na, Cs, and He${}^{\ast}$
interacting with a SiO${}_2$ plate (see the lines 1, 2, 3, and 4
in Fig.~3, respectively).
Taking into account that the influence of atomic properties is
more pronounced at the shortest separations, here we consider the
separation region from 100\,nm to $1\,\mu$m.
The atomic dynamic polarizabilities used in computations are
given by Eq.~(\ref{eq11}). For an atom of Na the oscillator
parameters are given by
$\alpha(0)=162.68\,$a.u., $\omega_0=2.14\,$eV \cite{48}.
For an atom of Cs one has
$\alpha(0)=403.6\,$a.u., $\omega_0=1.55\,$eV \cite{48,49},
and for an atom of He${}^{\ast}$ it holds
$\alpha(0)=315.638\,$a.u., $\omega_0=1.18\,$eV \cite{50}.
Note that the line 1 in Fig.~3 reproduces the initial part of
the upper line in Fig.~1.

As is seen in Fig.~3, the influence of graphene coating on the
free energy of the Casimir-Polder interaction increases with
decreasing characteristic frequency. {}From Fig.~3 it follows
also that at the shortest separations considered the ratio
${\cal F}_{\!\! g}/{\cal F}$ becomes nonmonotonous (this is most
appreciable for the Casimir-Polder interaction with an atom of
He${}^{\ast}$ marked by the maximum influence of graphene sheet).
This corresponds to the regions of negative Casimir-Polder and
Casimir entropy which were found in interactions of atoms
with metallic plates \cite{3} and between metallic and dielectric
plates \cite{51}.

\section{The Casimir-Polder force}

Although the Casimir-Polder free energy is the most important
quantity for the experiments on quantum reflection \cite{9,10,11},
the major role in the experiments on Bose-Einstein condensation
is played by the Casimir-Polder force. Here, we calculate this
force for the case of graphene-coated plates, find the role of
graphene, and obtain the analytic expressions for the classical
limit which holds at large separations (high temperatures).

\subsection{Influence of graphene coating on the force}

In terms of dimensionless variables introduced in Sec.~IIA
the Casimir-Polder force between an atom and a graphene-coated
plate is given by
\begin{eqnarray}
&&
F_g(a,T)=-\frac{k_BT}{8a^4}\sum_{l=0}^{\infty}
{\vphantom{sum}}^{\prime} \alpha(i\zeta_l\omega_c)
\int_{\zeta_l}^{\infty}\!\!\!ydye^{-y}
\label{eq12}\\
&&~
\times\left\{2y^2R_{\rm TM}(i\zeta_l,y)-\zeta_l^2\left[
R_{\rm TM}(i\zeta_l,y)+R_{\rm TE}(i\zeta_l,y)\right]
\right\},
\nonumber
\end{eqnarray}
\noindent
where the reflection coefficients $R_{\rm TM,TE}$ are defined
in Eq.~(\ref{eq5}).

To find the influence of graphene coating on the Casimir-Polder
force we have computed the ratio $F_g/F$ for the atom of
He${}^{\ast}$ at $T=300\,$K, where the index $g$ indicates that
the plate is coated with a graphene sheet. The computational
results, as functions of separation, are presented in Fig.~4 by
the four solid lines for Au, Si, Al${}_2$O${}_3$, and SiO${}_2$
plates. Similar to the free energy, it is seen that the
Casimir-Polder force between a He${}^{\ast}$ atom and an Au
plate is not influenced by the presence of graphene coating.
For dielectric plates the influence of graphene coating increases
with decreasing static dielectric permittivity of the plate
material. These results remain valid for any one of metallic and
dielectric plates. At short separation distances the influence
of graphene on the Casimir-Polder force is slightly larger than
on the free energy. Thus, at $a=100\,nm$, the quantity $F_g/F$
takes the values 1.014, 1.051, and 1.145 for Si, Al${}_2$O${}_3$,
and SiO${}_2$ plates, respectively. At $a=200\,nm$, the respective
values of $F_g/F$ are 1.018, 1.048, and 1.140, whereas at
$a=6\,\mu$m
they are eaual to 1.19, 1.22, and 1.70, i.e., the same as for the
free energy.

Similar to the case of the free energy, large influence of graphene
is explained by the extraordinary large thermal effect.
This can be illustrated in close analogy to Fig.~2(a,b)
presented above for the free energy. Similar to the free energy,
for different atoms the influence of graphene coating on the
Casimir-Polder force increases with decreasing characteristic
frequency of the atomic dynamic polarizability, like it is
shown in Fig.~3. Note that at short separations Fig.~4
demonstrates the nonmonotonous behavior which is explained
by the same reason as was discussed in connection with Fig.~3.

\subsection{High-temperature limit}

In the limiting case of high temperatures (large separations)
the main contribution to the Casimir-Polder and Casimir forces
becomes classical. It is well investigated for atoms interacting
with a graphene sheet \cite{34}, for two graphene sheets
\cite{28},
and for a graphene sheet interacting with a material plate
\cite{27}. Here, we consider the classical limit for an atom
interacting with a graphene-coated plate.

In the clasical limit the total Casimir-Polder free energy and
force are determined by the zero-frequency contribution to the
Lifshitz formula, whereas all other terms are exponentially
small \cite{40}. Thus, from Eq.~(\ref{eq1}) we have
\begin{equation}
{\cal F}_{\!\! g}(a,T)=-\frac{k_BT}{8a^3}\alpha(0)
\int_{0}^{\infty}\!\!\!y^2dye^{-y}R_{\rm TM}(0,y).
\label{eq13}
\end{equation}
\noindent
For the TM reflection coefficient at zero frequency
Eq.~(\ref{eq5}) leads to
\begin{eqnarray}
&&
R_{\rm TM}(0,y)=\frac{\varepsilon_0y-y+
\tilde{\Pi}_{00}(0,y)}{\varepsilon_0y+y+
\tilde{\Pi}_{00}(0,y)}
\label{eq14} \\
&&
\phantom{R_{\rm TM}(0,y)}=1-
\frac{2y}{\tilde{\Pi}_{00}(0,y)+(\varepsilon_0+1)y}.
\nonumber
\end{eqnarray}
\noindent
In the case of metallic plates $\varepsilon_0\to\infty$,
$R_{\rm TM}(0,y)\to 1$, and the classical limit (\ref{eq13})
takes the trivial form
\begin{equation}
{\cal F}_{\!\! g}(a,T)=-\frac{k_BT}{4a^3}\alpha(0),
\label{eq15}
\end{equation}
\noindent
which does not depend on the presence of graphene coating
in line with the above results. Below we consider a more
rich case of dielectric plates where $\varepsilon_0<\infty$.

We start from Eq.~(\ref{eq7}) and notice that in our region of
parameters (we consider the fixed temperature $T=300\,$K and
arbitrarily large separations $a\geq 100\,$nm) it holds
$\pi\theta/\tau\ll 1$. This is caused by the fact that the minimum
value of $\tau$
achieved at $a= 100\,$nm is equal to 0.164, and the main
contribution
to the integral (\ref{eq13}) is given by $y\sim 1$.
Then from Eq.~(\ref{eq7}) one obtains
\begin{equation}
\tilde{\Pi}_{00}(0,y)\equiv\tilde{\Pi}_{00}(0)\approx
\frac{8\alpha}{\tilde{v}_F^2}\,\frac{\tau}{\pi}\,\ln2
=32\alpha\ln2\frac{ak_BTc}{\hbar v_F^2}.
\label{eq16}
\end{equation}
\noindent
{}From Eq.~(\ref{eq16}) it follows that the quantity
$\tilde{\Pi}_{00}\sim 10^3$ and it increases with the
increase of separation. Thus, the first term in the
denominator of Eq.~(\ref{eq14}) (the second line) is much
larger than the second. Expanding the reflection coefficient
$R_{\rm TM}$ in power of small parameter, we arrive at
\begin{equation}
R_{\rm TM}(0,y)\approx 1-\frac{2y}{\tilde{\Pi}_{00}(0)}+
\frac{2(\varepsilon_0+1)}{\tilde{\Pi}_{00}^2(0)}y^2.
\label{eq17}
\end{equation}

Substituting Eq.~(\ref{eq17}) in Eq.~(\ref{eq13}) and
integrating with respect to $y$, we finally obtain
\begin{eqnarray}
&&
{\cal F}_{\!\! g}(a,T)\approx-\frac{k_BT\alpha(0)}{4a^3}\,
\left[\vphantom{\frac{(e_0)v_F^4}{8(k_BT)^2}}
1-\frac{3\hbar v_F^2}{16\alpha\ln2 ak_BTc}
\right.
\nonumber \\
&&~~~~~~~
\left.
+\frac{3(\varepsilon_0+1)\hbar^2v_F^4}{128(
\alpha\ln2 ak_BTc)^2}\right].
\label{eq18}
\end{eqnarray}
\noindent

Note that the first two terms on the right-hand side of
Eq.~(\ref{eq18}) coincide with the classical limit for the
Casimir-Polder free energy of an atom interacting with a
freestanding graphene sheet \cite{34}. The third term
on the right-hand side of Eq.~(\ref{eq18}) describes
the role of a dielectric plate. This term is small in
comparison with the first two. The main, classical, term
in Eq.~(\ref{eq18}) is the same as for an atom interacting
with metallic plate [see Eq.~(\ref{eq15})].

Equation~(\ref{eq18}) should be compared with the classical
Casimir-Polder free energy of an atom interacting with an
uncoated dielectric plate \cite{34}
\begin{equation}
{\cal F}(a,T)=-\frac{k_BT\alpha(0)}{4a^3}\,
\frac{\varepsilon_0-1}{\varepsilon_0+1}.
\label{eq19}
\end{equation}
\noindent
It is seen that Eq.~(\ref{eq19}) more strongly depends on the
static dielectric permittivity of plate material, as compared
to Eq.~(\ref{eq18}).

Similar derivations can be performed for the classical limit
of the Casimir-Polder force starting from Eq.~(\ref{eq12}).
Alternatively, the classical limit for this force can be
obtained by the negative differentiation of Eq.~(\ref{eq18})
with respect to $a$ with the following result:
\begin{eqnarray}
&&
{F}(a,T)\approx-\frac{3k_BT\alpha(0)}{4a^4}\,
\left[\vphantom{\frac{(e_0)v_F^4}{8(k_BT)^2}}
1-\frac{\hbar v_F^2}{4\alpha\ln2 ak_BTc}
\right.
\nonumber \\
&&~~~~~~~
\left.
+\frac{5(\varepsilon_0+1)\hbar^2v_F^4}{128(
\alpha\ln2 ak_BTc)^2}\right].
\label{eq20}
\end{eqnarray}
\noindent
The first two terms on the right-hand side of
Eq.~(\ref{eq20}) coincide with the
Casimir-Polder force acting between an atom and a
freestanding graphene sheet \cite{34}. The
influence of the dielectric plate is described
by the third term depending on $\varepsilon_0$.
The respective Casimir-Polder force for an atom
interacting with an uncoated dielectric plate is
given by \cite{34}
\begin{equation}
{F}(a,T)=-\frac{3k_BT\alpha(0)}{4a^4}\,
\frac{\varepsilon_0-1}{\varepsilon_0+1}.
\label{eq21}
\end{equation}
\noindent
The right-hand side of this equation more strongly depends
on $\varepsilon_0$ than the right-hand side of Eq.~(\ref{eq20}).
For a graphene-coated metallic plate the classical
Casimir-Polder force is the same as for an uncoated one
\begin{equation}
{F}_g(a,T)={F}(a,T)=-\frac{3k_BT\alpha(0)}{4a^4}.
\label{eq22}
\end{equation}

To determine the application region of Eqs.~(\ref{eq20}) and
(\ref{eq22}), we have performed numerical computations of the
Casimir-Polder forces $F_g$ at $T=300\,$K between an atom of
He${}^{\ast}$ and plates made of different materials coated
with graphene sheet. The computational results for the quantity
$|F_g|a^4$ as functions of separation are shown
in Fig.~5 by the four
solid lines from top to bottom for the plates made of Au,
Si, Al${}_2$O${}_3$, and SiO${}_2$, respectively.
In an inset the region of large separations is shown in an
enlarged scale. It can be seen that with increasing separation
distance the Casimir-Polder force between an atom of metastable
helium and graphene-coated plates made of different materials go
to one and the same limit given by the first term in
Eq.~(\ref{eq20}) and by Eq.~(\ref{eq22}).

We have compared the numerical results with the analytic ones
calculated using Eq.~(\ref{eq20}) for dielectrics and
Eq.~(\ref{eq22}) for metals. It was shown that for a
graphene-coated SiO${}_2$ plate (the smallest $\varepsilon_0$)
Eq.~(\ref{eq20}) leads to the same force values, as the computed
ones, in the limits of 2\% error at all separations
$a\geq 5\,\mu$m. For Al${}_2$O${}_3$ and Si plates the
coincidence between analytic and numerical results to
within 2\% occurs at $a\geq 5.5\,\mu$m.
As to the graphene-coated Au plate, the same measure of
agreement between the two sets of results holds at $a\geq 6\,\mu$m.

For comparison purposes in Fig.~6 we present the computational
results for the quantity $|F|a^4$, where $F$ is the
Casimir-Polder force at $T=300\,$K between an atom of
He${}^{\ast}$ and uncoated plates made of  Au,
Si, Al${}_2$O${}_3$, and SiO${}_2$ (the solid line is for Au,
and the dashed lines from top to bottom are for
Si, Al${}_2$O${}_3$, and SiO${}_2$, respectively).
The comparison of Figs.~6 and 5 shows that the upper solid
lines in both figures related to the Au plates coincide.
This is in line with the above results stating that graphene
coating on metallic plates does not influence the
Casimir-Polder interaction. As to the dashed lines in Fig.~6
related to Si, Al${}_2$O${}_3$, and SiO${}_2$ plates (from top
to bottom), they are significantly different from the
respective solid lines in Fig.~5. This is most pronounced
with increasing separations where the lines in Fig.~6 go to
the different limiting values prescribed by Eq.~(\ref{eq22})
for a metallic plate and by Eq.~(\ref{eq21}) for dielectric
plates with different $\varepsilon_0$.

The numerical results presented in Fig.~6 for an Au plate
(the solid line) coincide in the limits of 2\% error with
the analytic ones of Eq.~(\ref{eq22}) at $a\geq 6\,\mu$m.
For uncoated dielectric plates the numerical results
coincide with the analytic ones of Eq.~(\ref{eq21}) to
within 2\% starting from the same separations as for the plates
coated with graphene (i.e., at $a\geq 5\,\mu$m for SiO${}_2$
plate and at $a\geq 5.5\,\mu$m for Al${}_2$O${}_3$ and Si
plates). One can conclude that although the graphene coating of
dielectric plates significantly influences the Casimir-Polder
interaction, it does not change the minimum separation
distance starting with which the classical limit begins.
This is different from the case of atom-graphene Casimir-Polder
interaction where the classical limit occurs at $a\geq 1.5\,\mu$m
\cite{34}, i.e., starting from
by a factor of four shorter separations than
for an atom interacting with an Au plate.

\section{Conclusions and discussion}

In this paper we have calculated the Casimir-Polder free energies
and forces at $T=300\,$K between the ground state atoms of Rb,
Na, Cs and He${}^{\ast}$ and the plates made of Au, Si, sapphire
and fused silica coated with a graphene sheet. The obtained
results were compared with the free energies and forces for
uncoated plates, and the influence of graphene coating was
determined. It was found that the coating by graphene has no
effect on the Casimir-Polder free energy and force in the case
of metallic plates, but influences significantly for plates
made of dielectric materials. The impact of graphene coating was
shown to increase with decreasing static dielectric permittivity
of the plate material and characteristic frequency of the atomic
dynamic polarizability. Large influence of graphene coating on
the Casimir-Polder free energy and force was explained by the
extraordinary large thermal effect discovered earlier in the
van der Waals, Casimir and Casimir-Polder interactions with a
graphene sheet.

In the foregoing we have also obtained simple analytic
expressions for the classical limit of the Casimir-Polder free
energy and force between atoms and graphene-coated plates.
For metallic plates our result does not depend on the graphene
coating, but for dielectric plates is influenced by the
properties of graphene and (only slightly) by the static
dielectric permittivity of the plate material. With increasing
separation distance, the classical Casimir-Polder free
energy and force go to the quantities which do not depend on
the properties of graphene and material of the plate.
This is different from the case of uncoated dielectric plates
where the limiting values are material-dependent.
We have compared the analytic expressions for the classical
Casimir-Polder interaction with the results of numerical
computations and determined the application region of the
former.

The obtained results open prospective opportunities for using
the graphene-coated samples in experiments on quantum reflection,
Bose-Einstein condensation, and in various micro- and
nanodevices.


\begin{figure}[b]
\vspace*{-4cm}
\centerline{\hspace*{1cm}
\includegraphics{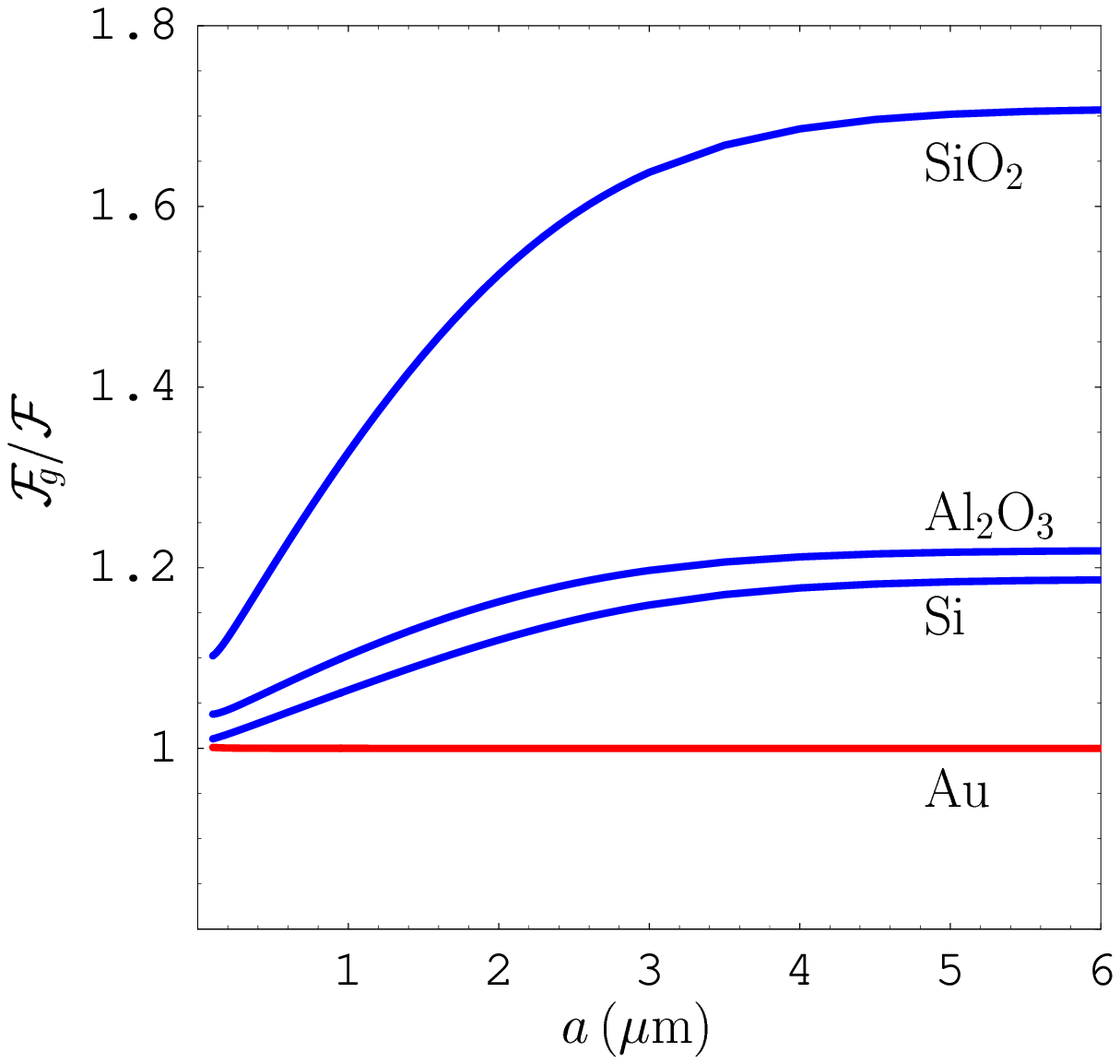}
}
\vspace*{-10cm}
\caption{\label{fg1}(Color online)
The ratios of the free energies of Casimir-Polder
interaction of a Rb atom with graphene-coated and
uncoated plates made of Au, Si, Al${}_2$O${}_3$,
and SiO${}_2$ are shown by the four solid lines
as functions of separation at $T=300\,$K.
}
\end{figure}
\begin{figure}[b]
\vspace*{-3cm}
\centerline{\hspace*{1cm}
\includegraphics{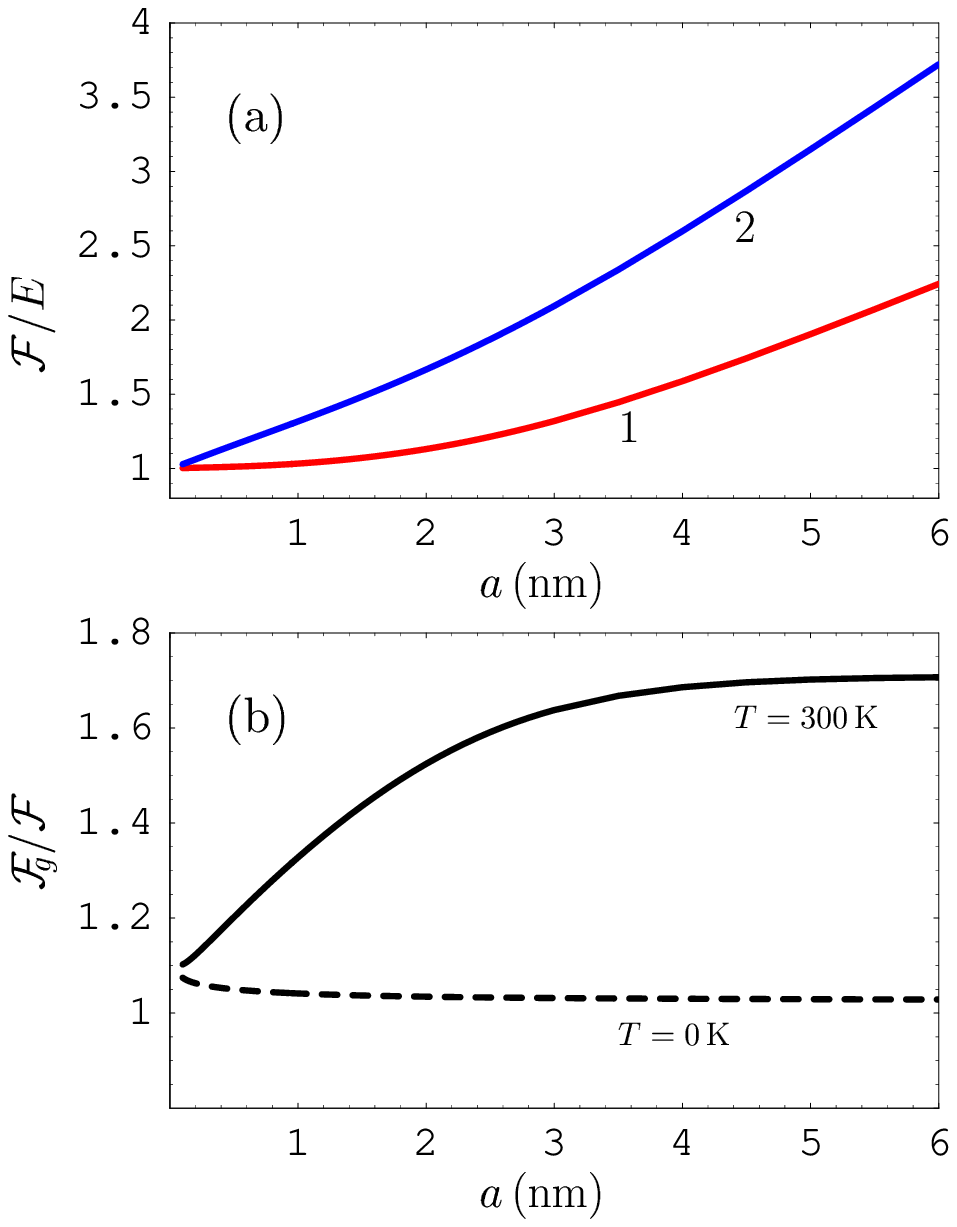}
}
\vspace*{-12cm}
\caption{\label{fg2}(Color online)
(a) The ratios of the free energy  at $T=300\,$K to the
energy  at $T=0\,$K for a Rb atom interacting with
an uncoated (the line 1) and graphene-coated (the line 2)
SiO${}_2$ plates are shown as functions of separation.
(b) The ratios of the free energies
of a Rb atom interacting with graphene-coated and
uncoated SiO${}_2$ plates
as functions of separation at $T=300\,$K (the solid
line) and  $T=0\,$K (the dashed line).
}
\end{figure}
\begin{figure}[b]
\vspace*{-4cm}
\centerline{\hspace*{1cm}
\includegraphics{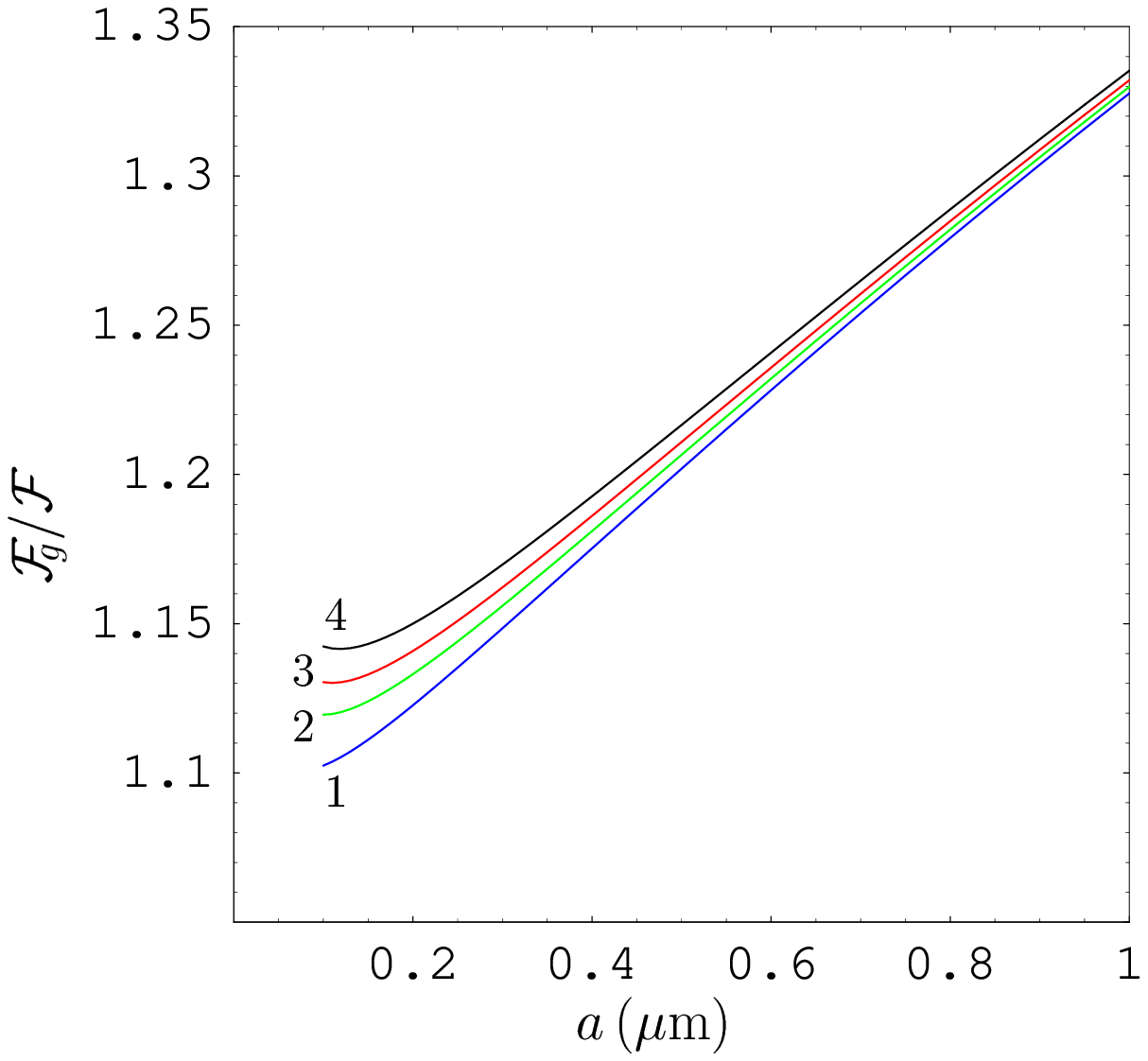}
}
\vspace*{-10cm}
\caption{\label{fg3}(Color online)
The ratios of the free energies of Casimir-Polder
interaction of a Rb, Na, Cs, and He${}^{\ast}$ atoms
with graphene-coated and
uncoated SiO${}_2$ plates are shown at $T=300\,$K
as functions of separation by the lines 1, 2, 3, and 4,
respectively.
}
\end{figure}
\begin{figure}[b]
\vspace*{-4cm}
\centerline{\hspace*{1cm}
\includegraphics{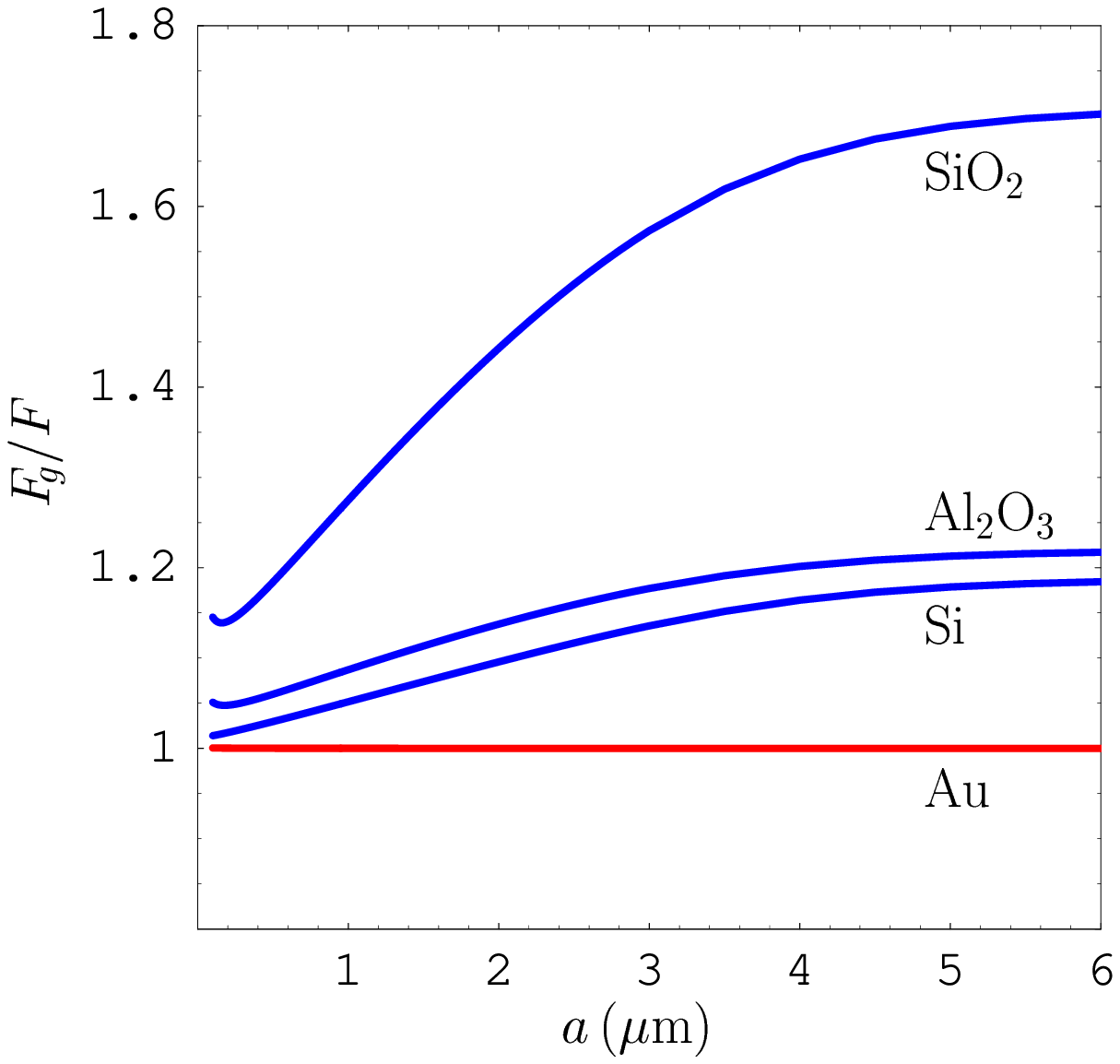}
}
\vspace*{-10cm}
\caption{\label{fg4}(Color online)
The ratios of the  Casimir-Polder forces between
a He${}^{\ast}$ atom and graphene-coated and
uncoated plates made of Au, Si, Al${}_2$O${}_3$,
and SiO${}_2$ are shown by the four solid lines
as functions of separation at $T=300\,$K.
}
\end{figure}
\begin{figure}[b]
\vspace*{-4cm}
\centerline{\hspace*{1cm}
\includegraphics{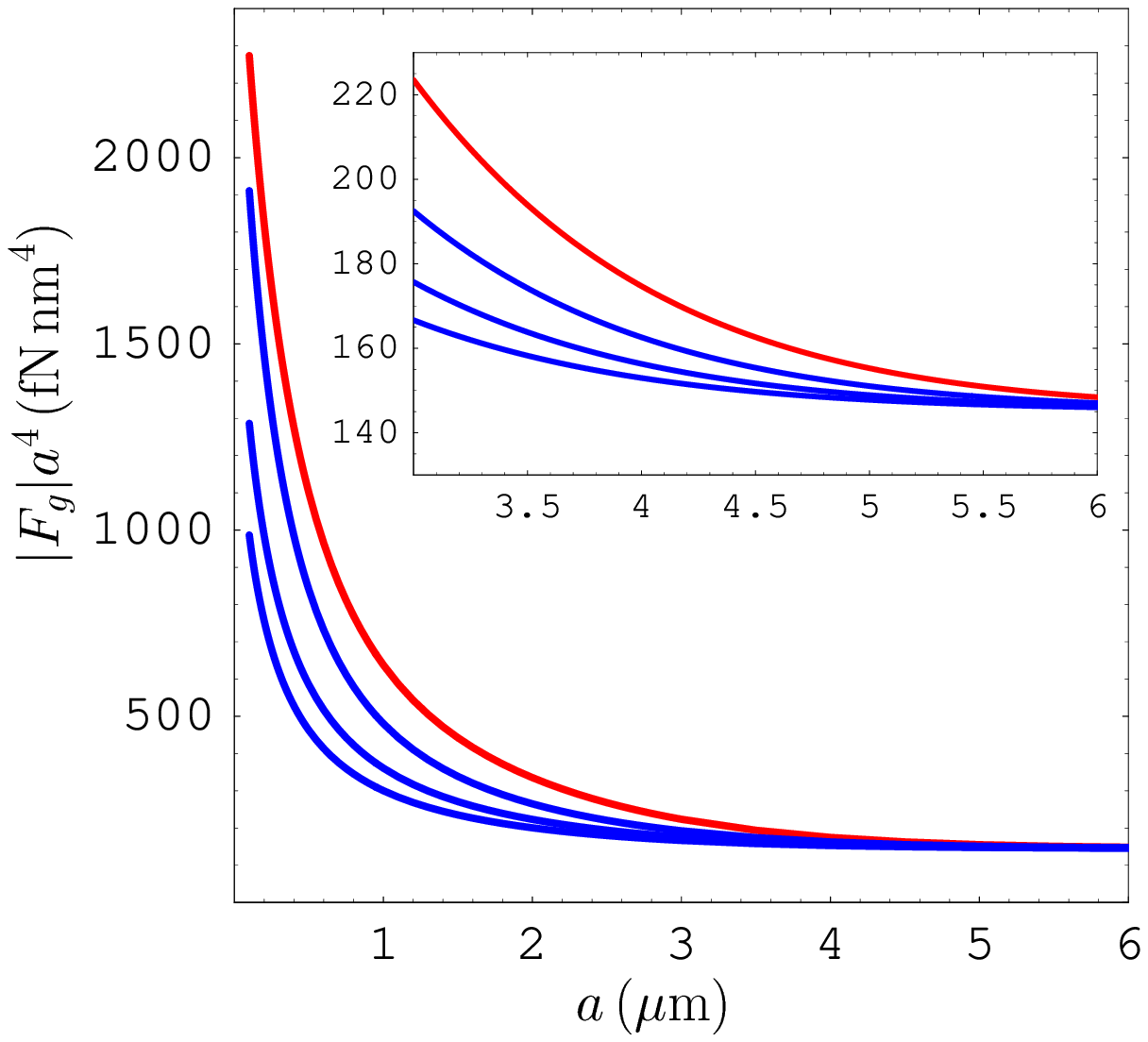}
}
\vspace*{-10cm}
\caption{\label{fg5}(Color online)
The magnitudes of the  Casimir-Polder force
multiplied by the fourth power of separation between
an atom of  He${}^{\ast}$ and graphene-coated
plates made of different materials at $T=300\,$K
are shown by the four solid lines from top to
bottom for the plates made of
Au, Si, Al${}_2$O${}_3$, and SiO${}_2$, respectively.
The inset shows the region of large separations on
an enlarged scale.
}
\end{figure}
\begin{figure}[b]
\vspace*{-4cm}
\centerline{\hspace*{1cm}
\includegraphics{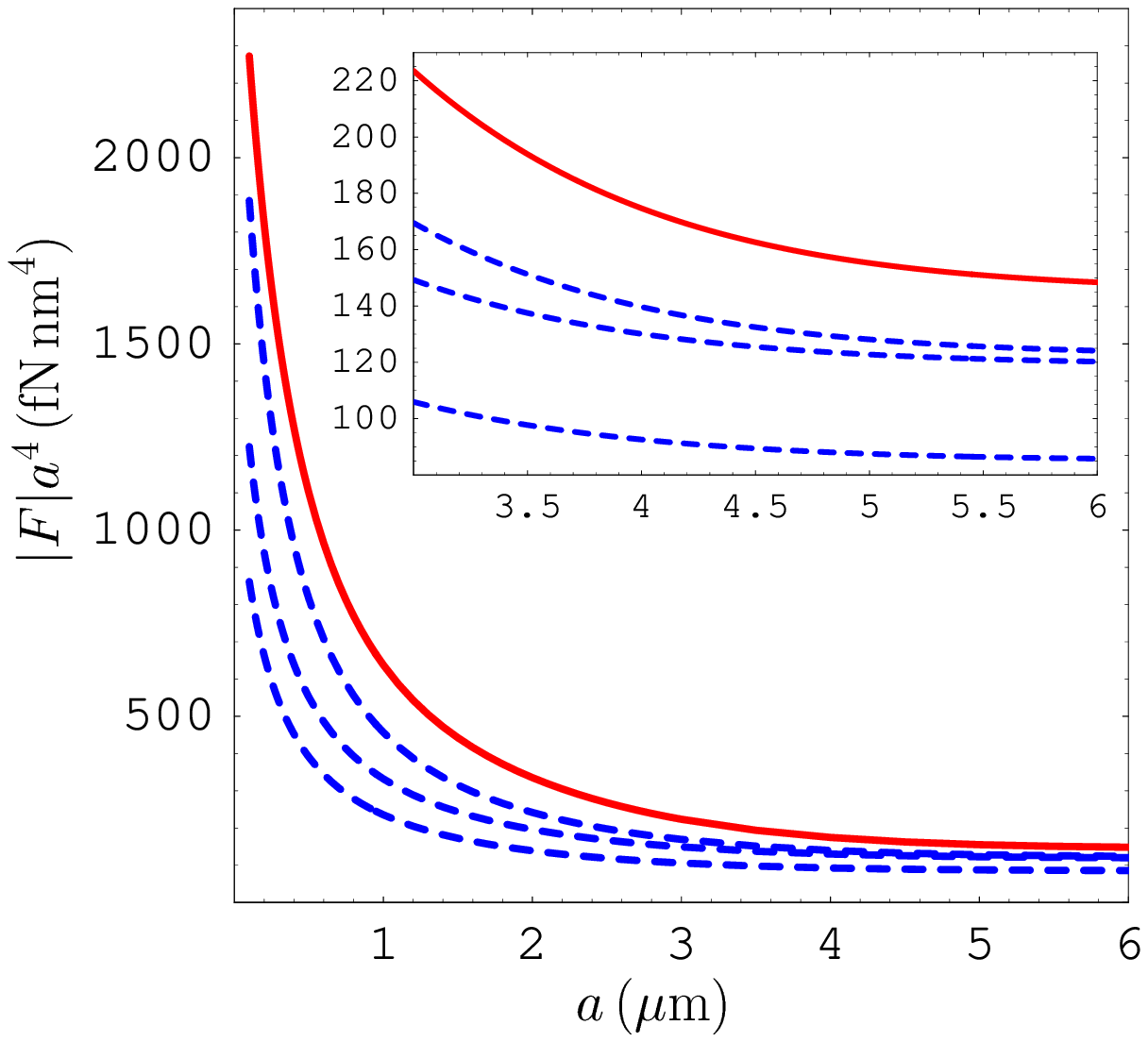}
}
\vspace*{-10cm}
\caption{\label{fg6}(Color online)
The magnitudes of the  Casimir-Polder force
multiplied by the fourth power of separation between
an atom of  He${}^{\ast}$ and uncoated
plates made of different materials at $T=300\,$K
are shown by the solid line for an Au plate and by the
three dashed lines from top to
bottom for the plates made of
 Si, Al${}_2$O${}_3$, and SiO${}_2$, respectively.
The inset shows the region of large separations on
an enlarged scale.
}
\end{figure}
\end{document}